\ifdefined\pdfminorversion\pdfminorversion=7\fi
\documentclass[journal]{IEEEtran}
\usepackage{amsmath,amsfonts}
\usepackage{algorithmic}
\usepackage{algorithm}
\usepackage{array}
\usepackage[caption=false,font=normalsize,labelfont=sf,textfont=sf]{subfig}
\usepackage{textcomp}
\usepackage{stfloats}
\usepackage{url}
\usepackage{verbatim}
\usepackage{graphicx}
\usepackage{epstopdf}
\usepackage{siunitx}
\usepackage{booktabs}
\usepackage[T1]{fontenc}
\usepackage{multirow}
\usepackage{diagbox}
\usepackage{balance}

\usepackage{hyperref}
\hypersetup{
  hidelinks,
  pdftitle={Smart-TCP: An Agentic AI-based Autonomous and Adaptive TCP Protocol},
  pdfauthor={Yule Han, Kezhi Wang, Yizhe Zhao, Kun Yang},
  pdfkeywords={Agentic AI, TCP, Large Language Models, Small Language Models, Network Service Management}
}
\begin{document}

\title{Smart-TCP: An Agentic AI-based Autonomous\\ and Adaptive TCP Protocol}

\author{Yule~Han, Kezhi~Wang,~\IEEEmembership{Senior~Member,~IEEE}, Yizhe~Zhao,~\IEEEmembership{Member,~IEEE}, and~Kun~Yang,~\IEEEmembership{Fellow,~IEEE}
\thanks{This work was supported in part by the National Natural Science Foundation of China (NSFC) under Grant No.62571091, Grant No.62431002, and Grant No.62132004, and in part by the Quzhou Government under Grant No.2024D007.}
\thanks{Yule Han is with the School of Information and Communication Engineering, University of Electronic Science and Technology of China, Chengdu 611731, China (e-mail: yulehan@std.uestc.edu.cn).}
\thanks{Kezhi Wang is with the Department of Computer Science, Brunel University London, UB8 3PH Uxbridge, U.K. (e-mail: kezhi.wang@brunel.ac.uk).}
\thanks{Yizhe Zhao is with the School of Information and Communication Engineering, University of Electronic Science and Technology of China, Chengdu 611731, China (e-mail: yzzhao@uestc.edu.cn).}
\thanks{Kun Yang is with the School of Information and Communication Engineering, University of Electronic Science and Technology of China, Chengdu, China, and also with the Institute of Intelligent Networks and Communications (NINE), Nanjing University, China (e-mail: kunyang@nju.edu.cn).}%
\thanks{Corresponding author: Kezhi Wang}
}


\maketitle

\begin{abstract}
The Transmission Control Protocol (TCP) relies on deterministic state machines and sequence-number arithmetic to ensure reliable communication. However, hard-coded protocol logic is difficult to adapt to increasingly complex and dynamic network conditions. This paper proposes Smart-TCP, an agentic AI-based transport protocol framework that organizes TCP control logic as a fast/slow model-assisted decision process. Specifically, a feature-aware classifier routes routine segments to a lightweight Small Language Model (SLM) fast path and anomalous or boundary cases to a Large Language Model (LLM) slow path, while an Arithmetic Logic Unit (ALU) handles deterministic computation of sequence and acknowledgment numbers. A state module maintains connection state, packet history, and key control variables, enabling protocol decisions over the full session context. A scoped Protocol Guard checks candidate decisions against RFC-derived rules before execution. We evaluate Smart-TCP through path classification, atomic packet processing, slow-path anomaly response, and end-to-end session interaction. Experimental results show that Smart-TCP achieves 99.14\% action accuracy on 700 slow-path anomaly and fallback samples, and a 100\% full-lifecycle success rate over 300 ideal fast-path end-to-end session windows. The ordinary 32-KiB transfer tests achieve a 96.5\% completion rate at 20\% segment loss. These results suggest that separating model-based protocol decisions from deterministic arithmetic improves the reliability of TCP control for network service operation.
\end{abstract}


\section{Introduction}
\label{sec:1}

The Transmission Control Protocol (TCP) remains the fundamental transport protocol for reliable Internet communication. Its control logic is specified as a deterministic state machine combined with sequence and acknowledgment arithmetic \cite{TCP-RFC}. This design enables interoperability across heterogeneous networks, but it also makes protocol behavior largely dependent on predefined rules. As mobile, wireless, and model-driven network environments become increasingly dynamic, transport protocols are expected to react not only to standard events, but also to heterogeneous access links, handoffs, anomalous packet orders, and context-dependent operational risks.

This pressure is becoming more visible in modern networked systems. Mobile users may move across radio access networks, middleboxes may rewrite or delay control segments, and heterogeneous paths may expose stale packets from previous connection contexts. At the same time, AI workloads introduce bursty and state-sensitive traffic patterns, making transport-layer behavior relevant to distributed training, inference serving, and multi-agent network simulation \cite{REASON24,RDMA-AI24}. In such environments, a segment that appears syntactically valid in isolation may still be semantically inconsistent with the current connection history. For example, an in-window reset segment may contradict a healthy bidirectional data-transfer history, while a zero-window advertisement may represent a transient receiver-side backpressure event, a lost window update, or an unrecoverable peer stall. These cases require reasoning over protocol state, packet fields, and historical context rather than only applying local header checks.

Existing research improves TCP, transport control, and service reliability from several perspectives. Learning-based transport designs such as TCP Ex Machina, PCC, and uncertainty-aware DRL congestion control show that sending behavior can be generated or adapted according to performance objectives, confidence estimates, and fallback policies rather than only fixed heuristics \cite{winstein2013tcp,PCC15,HAN_TNSM26}. Network-management studies also treat TCP behavior as part of service availability: edge-cloud defenses against TCP flood attacks protect service continuity under TCP-based DDoS traffic \cite{EdgeCookie_TNSM26}, while proxy-enabled networking architectures improve service delivery by delegating retry, routing, and protocol-translation functions to reconfigurable proxies \cite{Hermes_TNSM26}. These studies demonstrate that transport-layer behavior is increasingly coupled with network and service management objectives. Nevertheless, they mainly adapt congestion-control policies, attack-mitigation mechanisms, or service-delivery substrates, while the internal TCP state-transition and exception-handling loop remains largely implemented as a manually engineered finite-state machine.

In parallel, recent work on large language models (LLMs) for networking suggests a related direction. NetLLM studies how LLMs can be adapted to networking tasks \cite{NetLLM24}, while broader surveys discuss LLM-enabled wireless-network management and optimization \cite{WirelessLLMSurvey24}. Network foundation model studies explore foundation models for network-security tasks \cite{netFound23} and analyze the capabilities and limitations of such models \cite{DemystifyNFM24}. At the protocol level, ProtocolGPT uses LLMs to infer state machines from protocol implementations \cite{ProtocolGPT24}; a related LLM-agent study checks RFC conformance to detect functional bugs in protocol implementations \cite{RFCAudit25}; RFC-oriented parser validation translates specification text into traceable validation oracles \cite{RFCParserLLM25}; and LLM-guided protocol fuzzing uses language models to assist protocol testing \cite{LLMProtocolFuzzing24}. These studies indicate that language-model-based systems can capture useful protocol knowledge and support complex reasoning tasks that are difficult to encode with shallow rules alone. However, most existing LLM-assisted protocol studies still use the model as an external assistant: the model observes specifications, traces, or implementations and then produces analysis, test cases, or bug reports. The runtime protocol control logic itself is rarely placed under model-based decision-making.

Directly replacing TCP's control logic with a monolithic LLM is also not practical. First, routine segments require low-overhead processing, whereas large models introduce higher inference latency. Second, pure neural generation is unreliable for exact arithmetic on 32-bit sequence and acknowledgment numbers, which is central to TCP correctness. Third, complex boundary conditions require contextual reasoning, while routine data transfer should not pay the computational cost of large-model inference. These observations motivate separating context-dependent protocol decisions from deterministic arithmetic within a TCP research prototype.

Deterministic protocol checks complement exact sequence and acknowledgment arithmetic: they assess whether a model-selected action, next state, and control flags are permitted by the available connection state and packet fields. Smart-TCP therefore places an RFC-derived rule guard after model reasoning, with deterministic fallback for covered violations.

This paper therefore asks whether TCP control decisions can be organized as a model-assisted process with explicit connection memory and deterministic sequence-number arithmetic. Smart-TCP studies this question through a research prototype. Instead of treating language models as external tools for offline analysis, Smart-TCP embeds model-based reasoning into the protocol-control loop while explicitly separating routine processing, complex protocol-case reasoning, deterministic computation, and state memory.

Smart-TCP adopts a Dual-Brain architecture. A feature-aware classifier first extracts contextual features from the agent state, the received segment, and local action intent, and then routes the decision to either a fast or slow reasoning path. The fast path uses a lightweight Small Language Model (SLM) to handle high-frequency, low-risk, and structurally routine TCP decisions. The slow path uses a larger LLM to analyze anomalous or boundary cases, such as out-of-context control segments, malformed flag combinations, flow-control deadlocks, and benign packets that the classifier may misroute. In both paths, a deterministic Arithmetic Logic Unit (ALU) is invoked as a specialized tool to compute sequence and acknowledgment numbers exactly. A state module maintains connection memory, allowing the agent to reason over both current packet fields and historical interaction traces.

This decomposition follows the agentic AI paradigm of tool use and closed-loop execution \cite{sang2025beyond}, while avoiding a key weakness of pure LLM-based protocol generation: language models may reason about protocol context, but they should not be solely responsible for exact arithmetic \cite{baeumel2025lookahead}. In Smart-TCP, the language models select actions, states, flags, and tool calls, while the ALU provides verifiable arithmetic results.

Our contributions are summarized as follows:
\begin{itemize}
    \item We propose Smart-TCP, an agentic AI-based TCP control framework that organizes connection memory, model-assisted decisions, and deterministic tools in a research prototype.

    \item We design a feature-aware Dual-Brain architecture in which a classifier routes routine TCP decisions to a lightweight SLM and anomalous or boundary cases to an LLM, separating frequent packet handling from context-intensive protocol reasoning.

    \item We separate model-based protocol decisions from deterministic computation and safety checks: the ALU computes sequence and acknowledgment numbers, while a scoped RFC-derived guard checks candidate packet validity and whether proposed actions and next states are permitted in covered contexts, applying deterministic fallback when a covered constraint is violated.

    \item We evaluate classifier routing, fast-path packet generation, slow-path action prediction with a history ablation, prescribed ideal client--server session windows, and data transfer with segment loss, reordering, delayed ACKs, and concurrent flows. Constructed cases separately test the guard's covered rules.
\end{itemize}

\section{Related Work}
\label{sec:related_work}

\subsection{Adaptive Transport Control and Service Management}
Reliable network services increasingly require transport behavior to adapt to variable load, deployment constraints, and operational faults. Learning-based congestion-control studies, including PCC and HAN, show that transport protocols can improve throughput, latency, and robustness by adapting control policies to measured performance or model uncertainty \cite{PCC15,HAN_TNSM26}. TCP-aware service management has also been studied from the perspective of edge-cloud availability and service delivery: EdgeCookie mitigates TCP-based DDoS attacks in edge clouds \cite{EdgeCookie_TNSM26}, and Hermes delegates networking responsibilities to reconfigurable proxies to improve retry handling, routing, protocol translation, and service reliability \cite{Hermes_TNSM26}.

These studies show that transport behavior is a management concern for throughput, latency, service continuity, and edge-cloud availability. Nevertheless, they usually target specific mechanisms, such as congestion-window policy, TCP-flood mitigation, proxy-based retry/routing, or service-delivery architecture. Smart-TCP differs from them by exploring whether the protocol decision process itself can be organized as a model-assisted control loop, so that state transition reasoning, anomaly handling, and deterministic arithmetic are separated into specialized components.

At the wireless access layer, Guo et al. optimize beamforming and movable-antenna positions in a MISO system \cite{Guo2025FCLA}, while Lyu et al. jointly optimize power allocation, beamforming, and antenna positions in a movable-antenna-aided ISAC/NOMA system \cite{Lyu2025MAISAC}. As a cross-layer implication, variations in wireless capacity and reliability may affect packet loss, delay, and ACK timing observed by TCP. This possibility motivates examining TCP control decisions in the context of connection state and packet history; Smart-TCP does not model or optimize the physical layer.

\subsection{LLM-Assisted Networking and Protocol Engineering}
Recent work introduces large language models into networking and protocol engineering. NetLLM adapts LLMs for networking tasks, showing that language models can capture network-domain knowledge when properly tuned \cite{NetLLM24}. In network management, LCLE learns device-configuration sketches from manuals and command references, illustrating how LLMs can support operations-oriented automation \cite{LCLE_TNSM26}. Broader surveys discuss LLM-enabled wireless-network management and optimization \cite{WirelessLLMSurvey24}. Network foundation model studies further explore foundation models for network-security tasks \cite{netFound23} and analyze the capabilities and limitations of such models \cite{DemystifyNFM24}. Beyond management and foundation-model studies, ProtocolGPT uses LLMs to infer protocol state machines from protocol implementations, indicating that LLMs can assist in recovering protocol-level semantics \cite{ProtocolGPT24}. Other studies employ LLM-based agents or RFC interpretation to check RFC conformance and detect functional bugs in protocol implementations \cite{RFCAudit25}, translate RFC documents into traceable validation oracles for protocol parsers \cite{RFCParserLLM25}, and use language models to guide protocol fuzzing \cite{LLMProtocolFuzzing24}.

These works show that LLMs are useful tools for understanding, testing, and auditing network protocols. However, their role remains largely external: the model observes protocol specifications or implementations and then produces analysis, tests, or bug reports. Smart-TCP instead studies model-based decisions inside a prototype protocol-control loop. Smart-TCP does not ask the LLM or SLM to perform sequence-number arithmetic directly; instead, it couples model-based reasoning with a deterministic ALU, making the language model responsible for state and action selection while delegating exact arithmetic to a verifiable tool.

\begin{table*}[!t]
\caption{Positioning of Smart-TCP Relative to Representative Transport-Control and Network-Management Studies}
\label{tab:related_positioning}
\centering
\scriptsize
\setlength{\tabcolsep}{5pt}
\renewcommand{\arraystretch}{1.15}
\resizebox{\textwidth}{!}{%
\begin{tabular}{p{0.30\textwidth}cccc}
\toprule
\textbf{Line of Work} &
\textbf{Transport/Service Management} &
\textbf{Learning or LLM Method} &
\textbf{Protocol Semantic Reasoning} &
\textbf{Stateful Runtime TCP Control} \\
\midrule
TCP-aware service reliability management \cite{EdgeCookie_TNSM26,Hermes_TNSM26}
& $\surd$ & -- & -- & -- \\
\midrule
Learning-based transport control \cite{winstein2013tcp,PCC15,HAN_TNSM26}
& $\surd$ & $\surd$ & -- & -- \\
\midrule
LLM/network foundation models \cite{NetLLM24,WirelessLLMSurvey24,netFound23,DemystifyNFM24}
& -- & $\surd$ & -- & -- \\
\midrule
LLM-assisted protocol engineering \cite{ProtocolGPT24,RFCAudit25,RFCParserLLM25,LLMProtocolFuzzing24}
& -- & $\surd$ & $\surd$ & -- \\
\midrule
\textbf{Smart-TCP (Ours)}
& $\surd$ & $\surd$ & $\surd$ & $\surd$ \\
\bottomrule
\end{tabular}%
}
\end{table*}

Table~\ref{tab:related_positioning} summarizes the scope of the related lines of work. Existing transport-control and service-management studies mainly improve specific performance, reliability, or delivery mechanisms, while LLM-based protocol studies usually analyze or test protocols from outside the runtime data path. Smart-TCP differs by studying model-assisted protocol decisions in a stateful TCP control loop. The research prototype maintains connection state and packet history, while a deterministic ALU computes sequence and acknowledgment numbers when required.

\section{Agentic Framework Overview}
\label{sec:2}

This section describes the architecture and interaction of the client and server Smart-TCP agents. The framework uses a Large Language Model (LLM) and a lightweight Small Language Model (SLM) for TCP control decisions while keeping sequence-number arithmetic in a deterministic ALU. The goal is to study how hard-coded TCP control logic can be restructured around model-based decision making while preserving deterministic arithmetic execution.

\subsection{Design Principles and Scope}
\label{sec:scope}
Smart-TCP studies whether TCP control logic can be organized as a model-assisted runtime decision process. The language models propose protocol actions and state transitions, while deterministic modules compute sequence and acknowledgment numbers and perform protocol-safety checks.

Three design principles guide Smart-TCP. First, routine decisions follow a lightweight SLM fast path, while anomalous and boundary cases follow an LLM slow path that can use packet history. Second, model reasoning is separated from deterministic computation and protocol checks: the ALU computes sequence and acknowledgment numbers when required, and the integrated RFC-derived Protocol Guard checks candidate packet legality and whether proposed actions and next states are permitted within its covered rules, applying deterministic fallback for covered violations. Third, the state module maintains connection variables and packet history, making them available when a decision requires historical context.

\emph{Agentic} denotes coordination of the classifier, SLM/LLM, ALU, state module, and Protocol Guard; \emph{autonomous} denotes automated path selection and tool use within the evaluated interactions; and \emph{adaptive} denotes decisions conditioned on the packet, current state, and available history rather than online updates to the SFT-trained models. The implementation is a research prototype without an operating-system kernel, user-space TCP stack, or real socket interface integration.

\begin{figure*}[htbp]
    \centering
    \includegraphics[width=\linewidth]{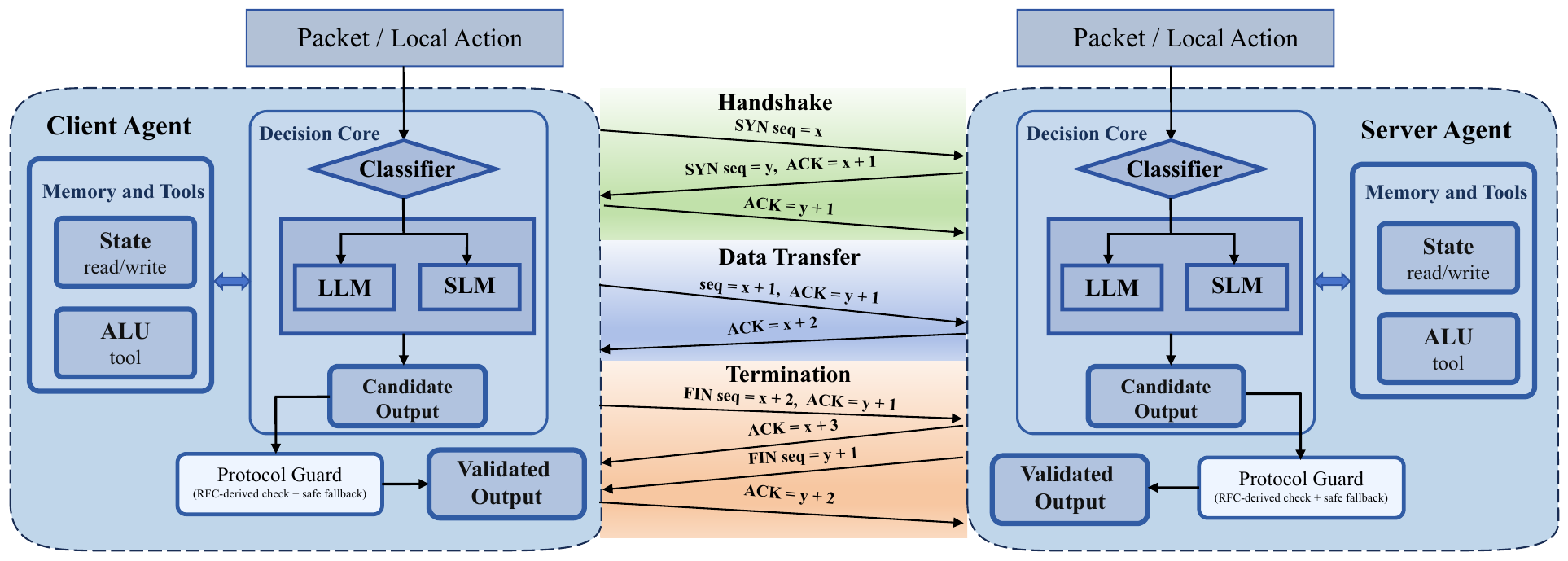}
    \caption{Dual-Brain Interaction Framework based on Smart-TCP Agents}
    \label{fig:1}
\end{figure*}

\subsection{\texorpdfstring{Dual-Agent Architecture}{Dual-Agent Architecture}}
Fig.~\ref{fig:1} shows a client agent and a server agent, each with its own Decision Core, Memory and Tools, and Protocol Guard. The ``Packet / Local Action'' input denotes a received segment or a locally triggered event, such as an application request or timer expiration. State and ALU are resources of the entire agent, rather than of either model path. The Decision Core produces a Candidate Output, which passes through the agent's Protocol Guard before reaching Validated Output. The guard's internal checks and fallback are detailed in Section~\ref{sec:protocol_guard}. The principal components are:

\begin{itemize}
    \item \textbf{Feature-Aware Classifier:} Acts as the initial classification gate for traffic. It analyzes current contextual features and determines whether the current protocol decision is routine or should be flagged for slow-path inspection.
    \item \textbf{Small Language Model (SLM / Fast Path):} Handles low-risk routine protocol procedures. It analyzes standard contexts, infers routine state transitions, and selects control flags with low processing overhead.
    \item \textbf{Large Language Model (LLM / Slow Path):} Handles anomalous or boundary protocol decisions. It infers appropriate handling methods when the classifier flags traffic such as out-of-order packets, flag conflicts, or ambiguous teardown-boundary states.
    \item \textbf{Arithmetic Logic Unit (ALU):} Calculates 32-bit sequence and acknowledgment numbers. Its role is limited to deterministic arithmetic execution.
    \item \textbf{State Module:} Stores the current connection state, including TCP phases and historical values, and records the updated protocol state after each decision.
    \item \textbf{Protocol Guard:} Checks candidate packet legality and whether proposed actions and next states are permitted by the implemented RFC-derived rules. It accepts a valid candidate under a covered rule, applies a checked deterministic replacement for a covered violation when available, or returns \texttt{unsupported} when no decision can be authorized.
\end{itemize}

\subsection{Protocol Lifecycle and Operational Loop}
\label{sec:operational_loop}
As shown in Fig.~\ref{fig:1}, the dual-brain agent interaction framework implements the complete TCP lifecycle. The interaction process is divided into the following three standard phases:

\begin{itemize}
    \item \textbf{Handshake:} The client and server agents establish a connection through the standard three-way handshake.
    \item \textbf{Data Transfer:} The agents exchange payload segments and acknowledgments after the connection enters the established state.
    \item \textbf{Termination:} The agents close the bidirectional connection through the standard four-segment termination procedure.
\end{itemize}

Under normal lifecycle execution, these routine segments are routed to the \textbf{SLM} fast path, while the ALU computes new sequence and acknowledgment numbers. The \textbf{LLM} slow path is invoked only when the Classifier identifies anomalous, boundary, or context-inconsistent protocol cases.

The reference architecture organizes each agent's decision path into four stages:

\begin{itemize}
    \item \textbf{Input and Routing:} The agent combines its State with the triggering packet or local action to form the decision context. The Classifier uses context features to select the SLM or LLM path.
    
    \item \textbf{Candidate Decision:} The selected model proposes a next state and packet control information; the LLM also proposes a handling action when selected. The agent's ALU performs any required arithmetic on 32-bit sequence and acknowledgment numbers during candidate preparation.
    
    \item \textbf{Protocol Check and Fallback:} The agent's Protocol Guard checks candidates from both paths against the trusted pre-decision context, including decisions that emit no segment, and applies deterministic fallback when needed. Only an accepted candidate or checked replacement reaches Validated Output. An \texttt{unsupported} result authorizes neither output nor a state update.

    \item \textbf{State Commit and Event Progression:} After authorization, the connection executor commits the decision and any state update, then processes the next queued event against the current connection state. Only one event per connection is active at a time.
\end{itemize}

During model inference, newly arriving segments and timer-expiration notifications enter the connection's event queue in occurrence order; enqueuing them does not modify its FSM or send/receive variables. The connection executor completes the active event before processing the next queued event against the current state. This policy applies to both model paths and serializes state updates within a connection, without requiring all connections to stop together.

Timer clocks continue during model inference. When a queued timer expiration is processed, the executor checks whether it is still valid under the current timer and connection state. For example, it ignores a queued retransmission timeout if the corresponding data have since been acknowledged or the timer has been canceled.

\section{Agent Implementation}
\label{sec:3}
In this section, we detail our implementation approach for the Smart-TCP agent, including: formalizing the tasks for the feature-aware classifier, the SLM/LLM decision modules, and the ALU; constructing the classification and protocol alignment datasets; designing unified interfaces; and implementing the training strategies for the multi-model pipeline.

\subsection{Formalization of Agent Tasks}
\label{sec:3_1}

We first formalize the feature-aware classifier ($\mathcal{L}_{Classifier}$), the dual-path decision models ($\mathcal{L}_{SLM}$ and $\mathcal{L}_{LLM}$), and the ALU ($\mathcal{L}_{ALU}$), followed by a separately tested protocol-safety interface. $S$ is the pre-decision connection memory, including the current TCP FSM state, send/receive control variables such as $SND.NXT$ and $RCV.NXT$, and relevant connection history. $R$ is the received TCP segment. $A$ is supplied local-event or processing-intent context, distinct from the model's output action $D_{act}$. These symbols denote information roles; their fields may be grouped within the implementation's state and event objects.

\subsubsection{\texorpdfstring{Feature-Aware Classifier ($\mathcal{L}_{Classifier}$)}{Feature-Aware Classifier}}
The classifier acts as the initial gating mechanism. It processes $S$, $R$, and $A$, and outputs a path-selection decision $p$:
$$p = \mathcal{L}_{Classifier}(S, R, A)$$
where $p=0$ indicates assigning the task to the SLM (fast path), and $p=1$ indicates assigning it to the LLM (slow path).

\subsubsection{\texorpdfstring{Dual-Path Reasoning Core ($\mathcal{L}_{SLM}$ and $\mathcal{L}_{LLM}$)}{Dual-Path Reasoning Core}}
Both models process $(S,R,A)$ and propose a next protocol state and associated memory update $S'$, TCP control flags $F$, a payload length $P_L$ in bytes, and an ALU task descriptor $T_{task}$. The slow-path LLM additionally proposes a handling directive $D_{act}$. A proposed state update is distinct from committing it to the state module.

When $p=0$, the task is handled by the SLM (fast path), and we formally define its reasoning process as:
\[(S', F, P_L, T_{task}) = \mathcal{L}_{SLM}(S, R, A)\]

When $p=1$, the task is handled by the LLM (slow path), and its reasoning process is defined as:
$$(D_{act}, S', F, P_L, T_{task}) = \mathcal{L}_{LLM}(S, R, A)$$

For the slow path, the action directive is defined as
\[
D_{act} \in \{\text{\texttt{FORWARD}}, \text{\texttt{DROP}}, \text{\texttt{RST}}, \text{\texttt{TRIGGER\_PROBE}}\}.
\]
\texttt{FORWARD} permits continued protocol processing; \texttt{DROP} discards the current input; \texttt{RST} requests a reset; and \texttt{TRIGGER\_PROBE} requests a zero-window probe. These are candidate handling directives. $T_{task}$ specifies deterministic arithmetic, such as \texttt{CALCULATE\_SEQ\_ACK}; \texttt{NONE} bypasses the ALU when no arithmetic is required.

\subsubsection{\texorpdfstring{Computational Tool ($\mathcal{L}_{ALU}$)}{Computational Tool}}
The $\mathcal{L}_{ALU}$ function performs deterministic computations when invoked. It calculates the Sequence ($Seq$) and Acknowledgment ($Ack$) numbers based on the task $T_{task}$, $S$, and $R$. We formalize the ALU's execution task as:
$$(Seq, Ack) = \mathcal{L}_{ALU}(T_{task}, S, R)$$

The outgoing segment $G$ is generated conditionally from the candidate decision and a separate emission indicator. \texttt{FORWARD} means that a valid input may continue through protocol processing; it does not require a segment to be emitted in the current step. When emission is authorized, the executor combines $(F,P_L)$ and any required ALU results $(Seq,Ack)$ to assemble $G$. A \texttt{DROP} decision emits no segment, while \texttt{RST} and \texttt{TRIGGER\_PROBE} require the state, sequence, and, for a probe, timer and retransmission conditions of the covered rule. Thus, when ALU arithmetic is required, its numerical correctness still does not validate the selected action or state.

\subsubsection{\texorpdfstring{Deterministic Protocol Check and Safe Fallback}{Deterministic Protocol Check and Safe Fallback}}
\label{sec:protocol_guard}
Let $X=(S,R,A)$ be the trusted pre-decision context. For the slow path, the action-stage checker reviews the LLM candidate $(D_{act},S')$ before packet assembly. The final checker reviews the structured candidate from either path, including the proposed action and next state, whether an outbound TCP segment is to be sent, and any associated packet fields or ALU task. The executor supplies $D_{act}$ for fast-path candidates; the SLM does not predict it. The checks use the current TCP state, numeric send/receive variables, the triggering segment or local event, and rule-specific timer or retransmission information. Model-generated history text is not treated as a trusted numeric predicate.

For each covered context, the checker compares the candidate with a finite set of allowed outputs; this set can contain more than one valid action or flag combination. When a covered violation has a defined fallback, the rejected candidate is replaced by a deterministic result derived from the same trusted context, without another SLM/LLM invocation. The replacement is structurally checked against the selected rule before the caller may commit it. If no rule or required field is available, or a rejected candidate has no valid fallback, the result is \texttt{unsupported}, with no authorization to emit the rejected candidate or apply its state update. The implementation returns \texttt{accepted}, \texttt{replaced}, or \texttt{unsupported} together with a structured packet decision and a proposed state patch when applicable.

The implemented rule subset covers simple handshake, in-order data, ordinary close, and selected reset and zero-window cases derived from RFC~9293~\cite{TCP-RFC}; reset handling follows the RFC~5961 protection branch described there.

\subsection{Dataset Engineering}
\label{sec:3_2}
To train the feature-aware classifier and the dual-path model component, we need state-labeled data that represents TCP interaction contexts. We design a two-stage data processing workflow to construct this multi-model learning environment:

\subsubsection{Data Sourcing and Path-Selection Labeling}
We source normal background traffic from the ISCX-VPN2016 dataset \cite{iscx}. By employing a 5-tuple filtering strategy and discarding incomplete connections, we extract 25,167 interaction records. These records cover the TCP lifecycle, from the three-way handshake to the four-way termination. To provide anomalous samples, we introduce the CIC-IDS-2017 dataset and construct a dataset of 13,778 records, primarily featuring out-of-order packets and flag tampering.

For Classifier training and evaluation, these 38,945 records are combined with 45,159 additional records sampled from the CIC-IDS-2017 Friday traffic. The reported Classifier result is evaluated on the held-out portion of this combined collection.

In the dual-brain architecture, these data labels serve specific classification and training purposes. The routine interaction records take $p=0$, while the anomalous records take $p=1$. These binary labels train the Classifier, equipping it with path-selection capability. Regarding the language models, the fast-path SLM uses normal TCP interaction data to learn routine lifecycle behavior, including handshake, data transfer, and termination. The LLM receives anomalous samples and benign fallback samples so that it can handle anomalies while still forwarding valid packets routed to the slow path.

\subsubsection{Retrospective State Reconstruction}
Standard network traces only record passive interaction outcomes. To enable supervised agent learning, we develop a custom retrospective analysis script to reconstruct the decision context behind each record. For each data record $G$ in the trace, the script performs a state-aware analysis of the flow history to restore the agent's memory state $S$ just before $G$ is generated. It also identifies the perceptual input $R$ triggering this response, while the local action $A$ is inferred directly from the content of $G$. This process transforms each $G$ into a complete \textit{(context, action)} sample pair, where the reconstructed $(S, R, A)$ tuple constitutes the model input.

We concurrently extract ground-truth behaviors from the raw trace data. In the current architecture, the reconstruction script outputs two levels of supervision signals: the first is the binary path-selection label $p \in \{0, 1\}$ dedicated to training the Classifier; the second is the behavior label for the language models, encompassing the next state $S'$, control flags $F$, and payload length $P_L$. Furthermore, we analyze TCP rules to derive the arithmetic task type $T_{task}$ required for $Seq$ and $Ack$ generation. Ultimately, the reconstructed $(S, R, A)$ serves as the model input, while $(S', F, P_L, T_{task})$ forms the behavioral guide for supervised fine-tuning of the SLM/LLM modules. The slow-path LLM has an additional action-directive target $D_{act}$.

\subsection{Agent Interfaces and Role Definition}
\label{sec:3_3}

Internal coordination within the Smart-TCP agent relies on structured interfaces. These interfaces separate high-level protocol decisions and exception handling from deterministic tool execution. This separation is achieved through the following three strategies:

\subsubsection{Structured Model Input}
The model input, the $(S, R, A)$ tuple (representing memory, perception, and action), is uniformly encoded as a structured JSON object. This step ensures that both the feature-aware Classifier and the backend model (SLM or LLM) can parse the multi-field context consistently. This standardized view supports contextual understanding and protocol decision-making.

\subsubsection{Decoupled Decision and Extended Action Space}
The model output is similarly formulated as a unified JSON object containing only protocol decisions. To accommodate both normal and anomalous scenarios, the interface is designed as an ``action superset'': it outputs fields determined by protocol logic (such as $S'$ and $F$) and a high-level tool invocation command $T_{task}$, alongside an extended \textit{Action Directive} (e.g., \texttt{FORWARD}, \texttt{DROP}, \texttt{RST}, \texttt{TRIGGER\_PROBE}).

For routine traffic, the agent outputs a \texttt{FORWARD} action accompanied by a valid $T_{task}$ (e.g., \texttt{CALCULATE\_SEQ\_ACK}), which is then delegated to the ALU for Sequence and Acknowledgment calculations. For anomalous traffic, when the LLM identifies invalid probes or protocol deviations, it can output a \texttt{DROP} or \texttt{RST} action, or issue \texttt{TRIGGER\_PROBE} to query the peer under flow-control deadlocks. When no outbound segment is required, $T_{task}$ is set to \texttt{NONE}, directly bypassing the ALU. The model never calculates $Seq$ or $Ack$ values directly; instead, it is responsible for issuing handling actions and tool scheduling rules.

The Protocol Guard checks the outputs of both paths against its implemented RFC-derived rules and the trusted connection context before any packet is sent or state update is committed.

\subsubsection{Path-Specific Prompting}
Both paths use English instructions. The fast-path instruction is:

\begin{quote}
\footnotesize
You are the Smart-TCP fast-path SLM. The classifier has routed the current event to the fast path. Given the connection state and send/receive control variables, the received TCP segment, and the local event or processing intent, return a structured candidate that specifies the next state, the necessary packet control information, and whether to invoke the ALU. If the ALU is needed, specify its computation task $T_{task}$. Do not compute $Seq$ or $Ack$ yourself, emit a packet, or commit a state update. The ALU performs the computation specified by $T_{task}$.
\end{quote}

The slow path considers the received segment, connection state, and recent sent/received packet history to infer the condition and propose a recovery action and next state. The following instruction is reproduced from the original 700-case slow-path evaluation records:

\begin{quote}
\footnotesize
You are the Smart-TCP Expert Control Plane. The mid-layer gating mechanism has forwarded a packet/event for inspection (numerical bounds have already been verified by hardware). Analyze the context, provide a root cause analysis, and specify the recovery action.
\end{quote}

The parenthetical hardware-verification statement is a supplied premise in this instruction, not evidence of an implemented hardware check. The software Protocol Guard is described in Section~\ref{sec:protocol_guard}. Table~\ref{tab:train_config} specifies the slow-path history length and both models' inference settings.

\subsection{Training the Decision Modules}
\label{sec:3_4}
We employ Supervised Fine-Tuning (SFT) to specialize generic neural network architectures for protocol-specific behavior. The training process encompasses both the routing gate and the SLM/LLM decision modules.

\begin{table}[htbp]
\centering
\caption{Slow-Path LLM Training and Inference Settings}
\label{tab:train_config}
\begingroup
\small
\setlength{\tabcolsep}{3pt}
\begin{tabular}{@{}p{0.38\columnwidth}p{0.55\columnwidth}@{}}
\toprule
\textbf{Parameter} & \textbf{Value} \\
\midrule
\multicolumn{2}{@{}l}{\textit{Slow-path LLM training}} \\
Model & Qwen2.5-7B \\
Fine-tuning method & LoRA \\
LoRA rank & 32 \\
Training precision & bf16 \\
Batch size & 2 \\
Grad. accum. steps & 2 \\
Epochs & 5 \\
Optimizer & AdamW \\
Learning rate & $1\times10^{-4}$ \\
Loss function & CrossEntropy \\
\midrule
\multicolumn{2}{@{}l}{\textit{Slow-path context}} \\
History length & Up to 5 recent sent/received packets \\
\midrule
\multicolumn{2}{@{}l}{\textit{Both paths' inference}} \\
Decoding & Greedy \\
\texttt{max\_new\_tokens} & 1024 \\
\bottomrule
\end{tabular}
\endgroup
\end{table}

\subsubsection{Gating and Base Model Strategy}
The classifier takes a $d$-dimensional feature vector and uses an MLP with three hidden layers; Table~\ref{tab:mlp_config} gives its network structure and training settings.

\begin{table}[htbp]
\centering
\caption{Classifier MLP Configuration}
\label{tab:mlp_config}
\begingroup
\small
\setlength{\tabcolsep}{3pt}
\begin{tabular}{@{}p{0.38\columnwidth}p{0.55\columnwidth}@{}}
\toprule
\textbf{Parameter} & \textbf{Value} \\
\midrule
Layer widths & $d\rightarrow128\rightarrow64\rightarrow32\rightarrow2$ \\
Activation & ReLU (hidden); softmax (output) \\
Dropout & 0.3 \\
Optimizer & Adam \\
Loss function & Sparse cross-entropy \\
Epochs & 20 \\
Batch size & 32 \\
\bottomrule
\end{tabular}
\endgroup
\end{table}

For the dual-path model component, we select models of different scales to match their respective roles: the lightweight \textbf{Qwen2.5-0.5B} serves as the fast path (SLM), while \textbf{Qwen2.5-7B} serves as the slow path (LLM). To achieve parameter-efficient fine-tuning, we employ Low-Rank Adaptation (LoRA) for both models, reducing the number of trainable parameters by freezing the pre-trained model weights. To preserve the LLM's capacity in the slow path to capture protocol anomaly features, we set its LoRA Rank to 32.

\subsubsection{Behavioral Alignment Objectives}
We use the native BPE tokenizer to process the model input $(S, R, A)$. The fine-tuning objective minimizes cross-entropy between each model's predicted structured output and its ground-truth target. For the SLM, the target is $(S', F, P_L, T_{task})$; for the LLM, it additionally includes $D_{act}$.

Both models learn to predict the next internal state, control flags, payload length, and ALU task; the LLM also learns the handling action. Arithmetic execution remains with the ALU.

Model training and fine-tuning run on a cloud server equipped with an NVIDIA A100 GPU. Table~\ref{tab:train_config} lists the slow-path LLM's training settings, the history length, and the decoding settings used by both selected models. The SLM adopts a similarly parameter-efficient configuration strategy.

\section{Experimental Evaluation}
\label{sec:4}

\subsection{Evaluation Questions}
To systematically evaluate the feasibility of Smart-TCP, the experiments are organized around four questions.

\textbf{RQ1:} Can the feature-aware classifier reliably distinguish routine TCP packets from complex or anomalous packets? This question evaluates the entry point of the dual-path architecture, because incorrect routing may either impose unnecessary large-model processing on benign traffic or allow abnormal packets to bypass slow-path diagnosis.

\textbf{RQ2:} Can the fast-path SLM accurately generate state transitions and control fields during routine TCP interactions? This question examines whether a lightweight model is sufficient for structured, high-frequency, and low-risk protocol decisions.

\textbf{RQ3:} Does separating sequence/acknowledgment arithmetic from model generation reduce cumulative errors in end-to-end session execution? This question tests the central design assumption that language models should reason about semantics and actions, while exact protocol arithmetic should be delegated to a deterministic ALU.

\textbf{RQ4:} Can the slow-path LLM use connection history to handle anomalies, boundary cases, and classifier false positives? This question evaluates whether the model can go beyond isolated field-level classification and reason jointly over FSM state, packet history, and protocol constraints.

\subsection{Experimental Setup and Baselines}
\label{sec:4_1}
To comprehensively evaluate the effectiveness of the multi-model Smart-TCP architecture (comprising the Classifier, SLM, LLM, and ALU), model training and fine-tuning run on a cloud server equipped with an NVIDIA A100 GPU, while single-packet inference latency benchmarks for the SLM and LLM decision modules run on a single NVIDIA RTX 5090 GPU. For the fast-path SLM, the 25,167 routine interaction records described in Section~\ref{sec:3_2} are randomly partitioned at the packet/record level with an 8:2 training/test ratio, without grouping by TCP connection. For the Classifier, the 38,945 traffic records described in Section~\ref{sec:3_2} are combined with 45,159 additional sampled records and randomly partitioned at the packet/record level with an 8:2 training/test ratio, without grouping by TCP connection. As described in Section~\ref{sec:3_2}, each reconstructed record represents a packet-level decision context rather than an entire connection; records from one connection may correspond to different TCP states and events. The held-out partition contains 16,821 records.

To facilitate systematic ablation and comparative analysis, we establish model and system baselines that correspond to the major components of Smart-TCP. For the fast path and end-to-end ideal-session tests, we compare the complete Smart-TCP design against SFT-only baselines such as \textbf{Qwen2.5-0.5B + SFT} and \textbf{Llama-3.2-1B + SFT}, which use the same structured protocol context but do not rely on the full classifier--ALU--state decomposition. For slow-path model selection, we compare Qwen2.5 models at different scales (1.5B, 3B, and 7B). These baselines isolate the effects of model scale, supervised protocol alignment, and deterministic arithmetic delegation.

Our evaluation metrics encompass both component-level and system-level dimensions. At the component level, we use classification accuracy to evaluate the Classifier, and we use accuracy and inference latency to evaluate the SLM and LLM decision modules. At the system level, Atomic Accuracy measures correctness across packet fields, End-to-End Trial Accuracy measures completion of the tested ideal session windows, and the data-transfer study reports completion, goodput, completion and recovery times, interruptions, and retransmissions.

\subsection{Evaluation of the Feature-Aware Classifier}
\label{sec:4_2}
Acting as the first line of defense in the dual-brain architecture, the feature-aware classifier is responsible for assigning traffic to either the fast path ($p=0$) or the slow path ($p=1$). This subsection evaluates the MLP with three hidden layers described in Section~\ref{sec:3_4}.

\begin{table}[htbp]
\centering
\caption{Classifier Confusion Matrix}
\label{tab:classifier_metrics}
\renewcommand{\arraystretch}{1.12}
\setlength{\tabcolsep}{6pt}
\begin{tabular}{@{}lcc@{}}
\toprule
\textbf{Reference path} & \multicolumn{2}{c}{\textbf{Predicted path}} \\
\cmidrule(l){2-3}
 & \textbf{Fast ($p=0$)} & \textbf{Slow ($p=1$)} \\
\midrule
Fast ($p=0$) & 8,853 & 6 \\
Slow ($p=1$) & 4 & 7,958 \\
\bottomrule
\end{tabular}
\end{table}

Table~\ref{tab:classifier_metrics} compares reference routing labels (rows) with Classifier predictions (columns). Routine records are labeled for the fast path ($p=0$), and anomalous records for the slow path ($p=1$). The false-positive rate (routine records sent to the slow path) is 0.0677\%, and the false-negative rate (anomalous records sent to the fast path) is 0.0502\%. Overall path-selection accuracy is 99.94\%, with an error rate of 0.06\%.

\subsection{Evaluating the Fast-Path SLM: Accuracy and Latency}
\label{sec:4_3}

This subsection evaluates the SLM-based fast path along two dimensions: protocol prediction accuracy and inference latency.

\subsubsection{Ablation Study on Protocol Precision}
To systematically dissect the factors contributing to protocol accuracy---specifically the impacts of model scale, domain-specific fine-tuning, and tool augmentation---we establish five distinct experimental configurations:
1) \textbf{Ours} (Qwen2.5-0.5B + SFT + ALU) represents our complete Smart-TCP fast-path agent; 
2) \textbf{M1} (Qwen2.5-0.5B + SFT) acts as an internal ablation to evaluate the base model's capability without the ALU; 
3) \textbf{M2} (Llama-3.2-1B + SFT) and 4) \textbf{M3} (Llama-3.1-8B + SFT) serve as external scale baselines to test if larger parameter counts can overcome arithmetic bottlenecks; 
5) \textbf{M4} (Llama-3.1-8B + Few-shot) represents the traditional paradigm of using general-purpose models without SFT.

\begin{figure}[htbp]
    \centering
    \includegraphics[width=\linewidth]{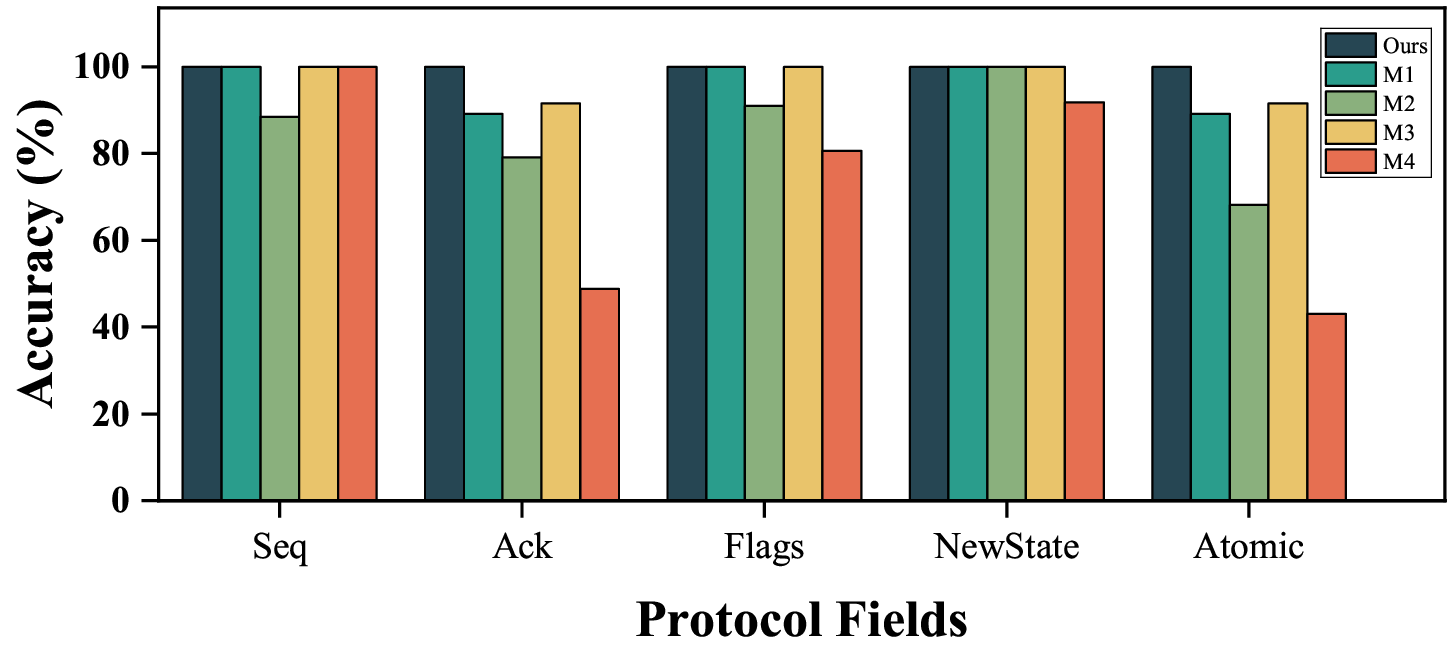}
    \caption{Fine-Grained Field-Level Accuracy. We compare our fast-path agent (\textbf{Ours}: Qwen2.5-0.5B + SFT + ALU) against four baselines: \textbf{M1} (Qwen2.5-0.5B + SFT), \textbf{M2} (Llama-3.2-1B + SFT), \textbf{M3} (Llama-3.1-8B + SFT), and \textbf{M4} (Llama-3.1-8B + Few-shot).}
    \label{fig:Fine-Grained Field-Level Accuracy}
\end{figure}

As shown in Fig.~\ref{fig:Fine-Grained Field-Level Accuracy}, the field-level accuracy comparison shows the effect of the decoupled architecture. For the \textit{NewState} field, all SFT-based configurations (Ours, M1, M2, and M3) achieve 100\% accuracy. For the \textit{Flags} field, Ours, M1, and M3 also reach 100\%, while M2 reaches 90.91\%. These results suggest that SFT substantially improves logical state-transition learning, outperforming the few-shot baseline M4 (80.61\% on \textit{Flags} and 91.82\% on \textit{NewState}). 

However, a performance gap appears in arithmetic-dependent fields. While most models perform well on the simpler \textit{Seq} field, pure neural networks show lower accuracy on the calculation-intensive \textit{Ack} field, regardless of domain-specific fine-tuning or parameter scale. Specifically, the few-shot baseline M4 reaches only 48.79\% on this field, and the SFT baselines M2, M1, and M3 reach 79.09\%, 89.09\%, and 91.52\%, respectively. In contrast, when the arithmetic task is delegated to the ALU, Smart-TCP (\textbf{Ours}) reaches 100\% accuracy. This indicates that arithmetic errors remain a limitation of pure LLM-based protocol generation in our experiments, even with SFT or parameter scaling.

We further evaluate the atomic accuracy, which requires simultaneous correctness across all protocol fields for a single packet. In this benchmark, Smart-TCP is the only configuration that achieves 100\% atomic accuracy. Notably, the atomic accuracies of M1 (89.09\%) and M3 (91.52\%) are exactly bottlenecked by their respective \textit{Ack} accuracies, while M2 is further affected by errors in \textit{Seq} and \textit{Flags}. This result supports combining a fine-tuned model with deterministic external tools for TCP packet generation.

\subsubsection{Inference Latency and Computational Efficiency}

Beyond protocol correctness, the practical value of the SLM path depends on reducing inference overhead relative to larger LLMs. In this prototype evaluation, the term ``fast path'' denotes a lower-latency model path compared with the 7B slow path, rather than a claim of deployment-ready data-plane throughput. All models in this comparison, including our SLM, use supervised fine-tuning (SFT) to align outputs with protocol logic. We utilize Flash Attention 2 for memory and compute optimization, and measure all latency benchmarks on a single NVIDIA RTX 5090 GPU.

\begin{table}[htbp]
\centering
\caption{Inference Latency and Atomic Accuracy Across Different Model Scales}
\label{tab:latency_results}
\begingroup
\renewcommand{\arraystretch}{1.2}
{ 
\renewcommand{\arraystretch}{1.2} 
\setlength{\tabcolsep}{4pt} 
\begin{tabular}{@{}lccc@{}}
\toprule
\multirow{2}{*}{\textbf{Model}} & \multirow{2}{*}{\textbf{Acc. (\%)}} & \multicolumn{2}{c}{\textbf{Latency (ms)}} \\ 
\cmidrule(l){3-4} 
 & & \textbf{Avg.} & \textbf{Min.} \\ 
\midrule
\textbf{Smart-TCP (Ours)} & \textbf{100.00} & 317.79 & 302.47 \\ 
\midrule
Qwen2.5-0.5B          & 89.09           & 314.79          & 301.63          \\
Llama-3.2-1B          & 68.18           & \textbf{287.38} & \textbf{259.18} \\
Llama-3.2-3B          & 87.58           & 572.92          & 553.46          \\
Llama-3.1-8B          & 91.52           & 1065.02         & 1020.77         \\ 
\bottomrule
\end{tabular}
} 
\endgroup
\end{table}

Table~\ref{tab:latency_results} presents the average and minimum inference latencies. Larger model scales generally introduce higher inference latency, although the exact value also depends on model architecture and implementation. The 8B model exhibits an average latency of 1065.02~ms (3.35$\times$ our approach), making it unsuitable for the low-latency path in this prototype despite an accuracy of 91.52\%. The Llama-3.2-1B model records the lowest average latency (287.38~ms); however, its atomic accuracy (68.18\%) fails to meet TCP state transition requirements. Our base model, Qwen2.5-0.5B, achieves 89.09\% accuracy with a latency of 314.79~ms.

Integrating the external ALU (\textbf{Ours}) introduces minimal computational overhead. Arithmetic execution and JSON parsing add approximately 3~ms to the processing time. Consequently, the Smart-TCP architecture maintains an average latency of 317.79~ms while achieving 100\% atomic accuracy. The design achieves deterministic arithmetic correctness on the evaluation set, which the larger pure neural baselines do not reach, while operating at a latency comparable to 1B-scale models.

\subsection{End-to-End Functional Validation}
\label{sec:4_4}

Section~\ref{sec:4_3} evaluates the fast-path model in static packet-level prediction. However, static accuracy does not guarantee successful dynamic interaction. To address this, we implement the dual-agent interaction framework illustrated in Fig.~\ref{fig:1}. Two Smart-TCP agents act as the Client and Server in ideal session tests covering the prescribed handshake, data-transfer, and termination sequence.

\subsubsection{Test Design}
\label{sec:4_4_1}

We execute 300 independent ideal session windows. Each test follows the prescribed handshake, data-transfer, and termination sequence. To isolate protocol execution correctness, this experiment introduces no stochastic packet loss, bit error, middlebox rewrite, or anomaly injection. All sessions utilize high-value random Initial Sequence Numbers ($ISN \in [2^{23}, 2^{32}-1]$). A phase counts as successful only when the agents reach the expected TCP FSM states and all outbound segments carry valid flags, payload lengths, and sequence/acknowledgment numbers. A trial counts as successful only if all three phases complete without protocol-state rollback or arithmetic inconsistency.

We compare the complete Smart-TCP agent against two SFT baselines from the fast-path evaluation: \textbf{Qwen2.5-0.5B + SFT} and \textbf{Llama-3.2-1B + SFT}. These baselines use the same structured protocol context but directly emit the protocol fields without the complete Smart-TCP decomposition, so they evaluate whether SFT alone can sustain closed-loop TCP communication.

\subsubsection{Results}
\label{sec:4_4_2}

\begin{table}[h!]
\centering
\caption{Phase-Level and End-to-End Results over 300 Ideal Session Windows}
\label{tab:e2e_comparison}
{
\renewcommand{\arraystretch}{1.08}
\setlength{\tabcolsep}{3pt}
\resizebox{\columnwidth}{!}{%
\begin{tabular}{l|ccc} 
\toprule
\textbf{Metric} & \textbf{Smart-TCP} & \textbf{Qwen2.5-0.5B + SFT} & \textbf{Llama-3.2-1B + SFT} \\ 
\midrule
Handshake       & \textbf{100.00\%} & 93.33\% & 96.67\% \\
Data Transfer   & \textbf{100.00\%} & 23.33\% & 3.33\% \\
Termination     & \textbf{100.00\%} & 96.67\% & 90.00\% \\ 
\midrule
\textbf{Trial Accuracy} & \textbf{100.00\%} & 23.33\% & 0.00\% \\
\bottomrule
\end{tabular}%
}
}
\end{table}

Table~\ref{tab:e2e_comparison} presents the end-to-end validation results. Smart-TCP achieves a trial accuracy of 100.00\% under the tested ideal session-window conditions. For the SFT-only baselines, the listed values are computed at the session-phase level: each 11-segment test window assigns the first three segments to handshake, the next four to data transfer, and the final four to connection termination. A phase is counted as successful only when all segments in that phase match the reference output. Under this stricter grouped-session criterion, Qwen2.5-0.5B + SFT reaches 23.33\% full-session accuracy, while Llama-3.2-1B + SFT does not complete any full test window. The dominant failure source is the data-transfer phase, where continuous acknowledgment-number updates amplify single-packet arithmetic errors.

This analysis supports the rationale behind the Smart-TCP design. In a stateful protocol like TCP, partial logical correctness is insufficient; an arithmetic inconsistency during data transfer can invalidate the entire session. Smart-TCP keeps protocol reasoning separate from packet arithmetic, which removes this failure mode under ideal end-to-end communication conditions.

\subsection{\texorpdfstring{Data Transfer with Loss, Reordering, and Delayed ACKs}{Data Transfer with Loss, Reordering, and Delayed ACKs}}
\label{sec:impaired_transfer}

We evaluate data-transfer completion using live model inference over a simulated link, varying segment loss, packet ordering, ACK timing, and flow count.

\subsubsection{\texorpdfstring{Test Design}{Test Design}}
We test ordinary delivery, reordering, delayed ACKs, and two concurrent connections at 0\%, 10\%, and 20\% segment loss, with 200 runs per condition and 32~KiB of data per connection. The Classifier and selected SLM/LLM make live inference calls. Initial transmissions and up to three timeout-triggered retransmissions use the same loss probability. We also test ordinary 128-KiB transfers with 200 runs each at 0\% and 20\% loss. For completed runs, goodput is the delivered payload divided by the time from the first send request before header generation to the final cumulative ACK; the table reports mean goodput and median completion and recovery times. Interruption and retransmission counts cover all runs in each 32-KiB condition.

\subsubsection{\texorpdfstring{Results}{Results}}

\begin{table*}[!t]
\caption{32-KiB Data-Transfer Results}
\label{tab:impaired_transfer}
\centering
\renewcommand{\arraystretch}{1.08}
\setlength{\tabcolsep}{4pt}
\resizebox{\textwidth}{!}{%
\begin{tabular}{llrrrrrr}
\toprule
\textbf{Transfer scenario} & \textbf{Loss} & \textbf{Completion rate} & \textbf{Goodput (KiB/s)} & \textbf{Completion (s)} & \textbf{Recovery (s)} & \textbf{Interrupted} & \textbf{Retransmissions} \\
\midrule
\multirow{3}{*}{Ordinary} & 0\% & 100.0\% & 1.46 & 21.85 & -- & 0 & 0 \\
 & 10\% & 100.0\% & 1.12 & 29.62 & 4.31 & 175 & 508 \\
 & 20\% & 96.5\% & 0.83 & 38.94 & 4.58 & 197 & 1142 \\
\midrule
\multirow{3}{*}{Reordering} & 0\% & 100.0\% & 1.43 & 22.48 & 1.93 & 12 & 0 \\
 & 10\% & 100.0\% & 1.10 & 28.45 & 3.98 & 180 & 526 \\
 & 20\% & 96.0\% & 0.82 & 39.20 & 4.39 & 197 & 1120 \\
\midrule
\multirow{3}{*}{Delayed ACK} & 0\% & 100.0\% & 1.52 & 21.60 & -- & 0 & 0 \\
 & 10\% & 99.5\% & 1.16 & 27.65 & 4.32 & 178 & 510 \\
 & 20\% & 98.0\% & 0.86 & 39.15 & 4.52 & 199 & 1085 \\
\midrule
\multirow{3}{*}{Two concurrent flows} & 0\% & 99.5\% & 1.48 & 43.20 & -- & 16 & 21 \\
 & 10\% & 100.0\% & 1.37 & 46.10 & 5.85 & 198 & 968 \\
 & 20\% & 94.0\% & 1.16 & 54.75 & 5.88 & 200 & 2218 \\
\bottomrule
\end{tabular}%
}
\end{table*}

Table~\ref{tab:impaired_transfer} shows that completion remains at or above 99.5\% across the four 32-KiB conditions at 0\% loss and ranges from 94.0\% to 98.0\% at 20\% loss. At 20\% loss, the completion rates are 96.5\% for ordinary delivery, 96.0\% with reordering, 98.0\% with delayed ACKs, and 94.0\% for two concurrent flows. For ordinary delivery, mean goodput decreases from 1.46 to 0.83~KiB/s as loss rises from 0\% to 20\%, while median completion time increases from 21.85 to 38.94~s. The two-flow condition has the longest median completion time (54.75~s) and recovery time (5.88~s) at 20\% loss. Interruption and retransmission totals also rise with loss across all four conditions.

\begin{table}[!t]
\centering
\caption{Ordinary-Transfer Completion Rates by Data Size}
\label{tab:extended_transfer}
\small
\setlength{\tabcolsep}{10pt}
\renewcommand{\arraystretch}{1.18}
\begin{tabular}{@{}ccc@{}}
\toprule
\textbf{Segment loss} & \textbf{32 KiB} & \textbf{128 KiB} \\
\midrule
0\% & 100.0\% & 98.0\% \\
20\% & 96.5\% & 80.5\% \\
\bottomrule
\end{tabular}
\end{table}

Table~\ref{tab:extended_transfer} compares ordinary-transfer completion at 32~KiB and 128~KiB; its 32-KiB values come from Table~\ref{tab:impaired_transfer}. As segment loss rises from 0\% to 20\%, the completion rate decreases from 100.0\% to 96.5\% for 32~KiB and from 98.0\% to 80.5\% for 128~KiB. The longer transfer is more affected by loss under the tested conditions.

\subsection{Slow Path: LLM Handling of Protocol Anomalies and Fallback Mechanism}
\label{subsec:slow_path}

The preceding sections evaluate the fast-path components and ideal-session end-to-end behavior: the Feature-Aware Classifier achieves accurate path selection, while the SLM + ALU fast path generates state transitions and sequence/acknowledgment updates under benign operational scenarios. However, real-world TCP sessions do not always follow structured and conventional execution paths. In environments involving mobile handovers, NAT rebinding, multi-path routing changes, middlebox rewrites, lost window updates, and four-way termination boundaries, protocol control segments may appear syntactically valid on a per-field basis but be inconsistent with the current connection history. Such complex protocol cases cannot be resolved merely by evaluating static header fields or deploying shallow binary classifiers; rather, they require the system to reason over historical context, current finite state machine (FSM) states, and behavioral constraints mandated by RFC 9293.

Consequently, this section evaluates the slow-path LLM Agent. In contrast to the fast-path SLM in Section~\ref{sec:4_3}, which primarily handles regular traffic ($p=0$), the slow path is dedicated to anomalous, complex, or boundary segments flagged by the Classifier ($p=1$). We evaluate whether the LLM can diagnose anomalies and predict candidate actions, including \texttt{DROP}, \texttt{RST}, \texttt{TRIGGER\_PROBE}, and \texttt{FORWARD}.

Our empirical evaluation is structured around three questions:
\begin{enumerate}
    \item \textit{Diagnosis Capability for Out-of-Context Control Segments}: Can the LLM use connection history to identify suspicious control segments and propose an appropriate action for subsequent protocol checking?
    \item \textit{Stateful Triage for Flow-Control Deadlocks}: When multiple heterogeneous sub-scenarios share the identical surface-level feature of a zero-window ($\text{Win}=0$), can the LLM perform fine-grained root-cause analysis and execute differentiated recovery strategies?
    \item \textit{Handling Classifier False Positives}: When the upstream Classifier occasionally misclassifies benign traffic as anomalous, can the LLM use the packet, state, and history to identify valid traffic and propose a \texttt{FORWARD} action?
\end{enumerate}

\subsubsection{Setup of the Anomaly-Aware SFT Dataset and Base Model Selection}
To evaluate the protocol reasoning capabilities of the slow-path LLM, we use two sources of supervision. First, we reuse the normal TCP interaction dataset reconstructed from realistic traffic traces. This source teaches the LLM to forward valid handshake, data-transfer, and termination packets when the Classifier mistakenly routes them to the slow path. Second, we construct an anomaly-aware SFT dataset, $D_{\text{new}}$, containing 8,000 training samples and 700 evaluation samples across 16 predefined protocol sub-scenarios. The dataset covers out-of-context control segments, stateful misalignment/flow-control cases, and benign traffic misrouted by the classifier.

Specifically, $D_{\text{new}}$ comprises 16 representative protocol-boundary sub-scenarios, condensed into three categories: out-of-context diagnostic segments, stateful misalignment/flow-control deadlocks, and benign fallback traffic. The out-of-context diagnostic category includes both residual reset events and malformed control-flag combinations, thereby matching the integrated evaluation in Table~\ref{tab:comprehensive_slow_path}. All samples are formatted into \texttt{(instruction, input, output)} triplets. The \texttt{input} field includes the current TCP state, packet headers, and historical context. The reference output for each evaluation record contains root-cause text, a recovery-action label, and the next FSM state; the recovery-action label is used to assess the LLM's proposed action. The overall data distribution is summarized in Table~\ref{tab:sft_dataset}.

\begin{table}[htbp]
\caption{Composition of the Anomaly-Aware SFT Dataset ($D_{\text{new}}$) for the Slow Path}
\label{tab:sft_dataset}
\centering
\scriptsize
\setlength{\tabcolsep}{2.5pt}
\renewcommand{\arraystretch}{1.12}
\resizebox{\columnwidth}{!}{%
\begin{tabular}{lcccc}
\toprule
\textbf{Scenario Category} & \textbf{Sub-scenarios} & \textbf{Train} & \textbf{Eval} & \textbf{Dataset Action Label} \\ 
\midrule
Out-of-Context Diagnostics  & 4                      & 2200                   & 200                   & \texttt{DROP} \\
Stateful Misalignment       & 5                      & 2400                   & 200                   & \texttt{TRIGGER\_PROBE/RST/DROP} \\
Benign Fallback Traffic     & 7                      & 3400                   & 300                   & \texttt{FORWARD} \\ 
\midrule
\textbf{Total}              & \textbf{16}            & \textbf{8000}          & \textbf{700}          & \textit{--} \\ 
\bottomrule
\end{tabular}%
}
\end{table}

\paragraph*{Slow-Path Base Model Ablation and Selection}
To select the slow-path model, we conduct an ablation study evaluating the Qwen2.5 series across different parameter scales (1.5B, 3B, and 7B). All ablation experiments run on a single NVIDIA RTX 5090 GPU. The evaluation benchmark is the 700-sample slow-path test set. The number of decisions matching the original action labels, label agreement, GPU memory footprint, and inference latency statistics computed over all 700 evaluation samples for each model are reported in Table~\ref{tab:model_selection}.

\begin{table*}[!t]
\caption{Candidate LLM Agreement with the Original 700 Slow-Path Action Labels and Inference Latency on RTX 5090}
\label{tab:model_selection}
\centering
\scriptsize
\setlength{\tabcolsep}{4pt}
\renewcommand{\arraystretch}{1.12}
\begin{tabular*}{\textwidth}{@{\extracolsep{\fill}}lcccccc@{}}
\toprule
\multirow{2}{*}{\textbf{Model}} & \multirow{2}{*}{\textbf{Match/Total}} & \multirow{2}{*}{\textbf{Label Agr.}} & \multirow{2}{*}{\textbf{GPU Mem. (GB)}} & \multicolumn{3}{c}{\textbf{Latency (ms)}} \\
\cmidrule(l){5-7}
 & & & & \textbf{Avg.} & \textbf{Min.} & \textbf{Max.} \\ 
\midrule
Qwen2.5-1.5B                         & 639/700                  & 91.29\%                         & 3.39                     & 420.91 & 393.23 & 444.98 \\
Qwen2.5-3B                           & 674/700                  & 96.29\%                         & 6.41                     & 618.68 & 594.05 & 646.20 \\
\textbf{Qwen2.5-7B (Ours)}           & \textbf{694/700}         & \textbf{99.14\%}                & \textbf{14.82}           & \textbf{1033.84} & \textbf{855.59} & \textbf{1320.51} \\ 
\bottomrule
\end{tabular*}
\end{table*}

Agreement with the dataset's action labels rises from 91.29\% (639/700) for 1.5B to 96.29\% (674/700) for 3B and 99.14\% (694/700) for 7B. The 7B model has a reported average single-packet latency of 1033.84~ms and is selected as the slow-path model for this evaluation.

The latency values in Table~\ref{tab:model_selection} summarize inference over all 700 evaluation samples for each candidate model. The table reports the mean, minimum, and maximum observed latency under this test setting.

We evaluate the selected slow-path model on the same 700 reference cases using three inputs: packet only, packet plus current state, and packet plus state and historical context. Recovery-action accuracy counts both the chosen handling and its associated state correction. Table~\ref{tab:comprehensive_slow_path} compares the three conditions.
\begin{table*}[!t]
\caption{Slow-Path Recovery-Action Accuracy Across 16 Sub-scenarios and Three Input Conditions}
\label{tab:comprehensive_slow_path}
\centering
\scriptsize
\setlength{\tabcolsep}{2pt}
\renewcommand{\arraystretch}{1.08}
\resizebox{\textwidth}{!}{%
\begin{tabular}{@{}llccccc@{}}
\toprule
\textbf{Evaluation Dimension} & \textbf{Protocol Sub-scenario / FSM Phase} & \textbf{N} & \textbf{Dataset Label} & \textbf{Packet Only} & \textbf{Packet+State} & \textbf{Packet+State+History} \\ 
\midrule
\textbf{1) Out-of-Context Diagnostics}  & Residual RST in Data Transfer Phase                & 50               & \texttt{DROP}                     & 58.0\% & 16.0\% & 100.0\% \\
                                        & Out-of-Context RST in \texttt{FIN\_WAIT} States    & 50               & \texttt{DROP}                     & 0.0\% & 80.0\% & 98.0\% \\
                                        & Isolated FIN with Missing ACK (\textit{Bare FIN}) & 50               & \texttt{DROP}                     & 0.0\% & 0.0\% & 100.0\% \\
                                        & Malformed \texttt{SYN+PSH} Segment                 & 50               & \texttt{DROP}                     & 0.0\% & 0.0\% & 100.0\% \\
\cmidrule{2-7}
                                        & \textit{Category Subtotal / Average}               & \textit{200}     & \textit{--}                      & \textit{14.50\%} & \textit{24.00\%} & \textit{99.50\%} \\
\midrule
\textbf{2) Stateful Deadlock Triage}    & Gradual Window Exhaustion                          & 40               & \texttt{TRIGGER\_PROBE}           & 0.0\% & 0.0\% & 100.0\% \\
                                        & Abrupt Window Zeroing                              & 40               & \texttt{TRIGGER\_PROBE}           & 0.0\% & 0.0\% & 97.5\% \\
                                        & Prolonged Window Deadlock                          & 40               & \texttt{TRIGGER\_PROBE}           & 0.0\% & 0.0\% & 100.0\% \\
                                        & Zero-window Probe Unresponsiveness                 & 40               & \texttt{RST}                      & 0.0\% & 0.0\% & 97.5\% \\
                                        & Illegal Window Shrinkage                           & 40               & \texttt{DROP}                     & 0.0\% & 0.0\% & 100.0\% \\
\cmidrule{2-7}
                                        & \textit{Category Subtotal / Average}               & \textit{200}     & \textit{--}                      & \textit{0.00\%} & \textit{0.00\%} & \textit{99.00\%} \\
\midrule
\textbf{3) Fallback Interoperability}   & \texttt{SYN\_SENT} Valid SYN-ACK Response          & 40               & \texttt{FORWARD}                  & 0.0\% & 100.0\% & 100.0\% \\
                                        & \texttt{ESTABLISHED} Conventional Data Segment     & 50               & \texttt{FORWARD}                  & 0.0\% & 100.0\% & 100.0\% \\
                                        & \texttt{FIN\_WAIT\_1} Legitimate ACK Reception     & 40               & \texttt{FORWARD}                  & 0.0\% & 2.5\% & 100.0\% \\
                                        & \texttt{FIN\_WAIT\_2} Peer FIN Arrival             & 40               & \texttt{FORWARD}                  & 0.0\% & 40.0\% & 97.5\% \\
                                        & \texttt{CLOSE\_WAIT} Passive Close Legitimate ACK  & 40               & \texttt{FORWARD}                  & 0.0\% & 0.0\% & 97.5\% \\
                                        & \texttt{LAST\_ACK} Complex Retransmission Boundary & 50               & \texttt{FORWARD}                  & 0.0\% & 70.0\% & 98.0\% \\
                                        & \texttt{TIME\_WAIT} Protective FIN Retransmission  & 40               & \texttt{FORWARD}                  & 0.0\% & 67.5\% & 100.0\% \\
\cmidrule{2-7}
                                        & \textit{Category Subtotal / Average}               & \textit{300}     & \textit{--}                      & \textit{0.00\%} & \textit{56.33\%} & \textit{99.00\%} \\
\midrule
\multicolumn{2}{l}{\textbf{Overall Recovery-Action Accuracy}}                           & \textbf{700}     & \textbf{--}                      & \textbf{4.14\%} & \textbf{31.00\%} & \textbf{99.14\%} \\
\bottomrule
\end{tabular}%
}
\end{table*}

The overall recovery-action accuracies are 29/700 (4.14\%) with packet only, 217/700 (31.00\%) with packet and state, and 694/700 (99.14\%) with packet, state, and historical context. The six original full-input non-matches comprise one reset case, two zero-window cases, and three teardown-adjacent benign cases.

\subsubsection{Diagnosis Capability for Out-of-Context Control Segments}
We first evaluate the LLM's diagnostic capability when encountering out-of-context control segments. In real-world networks, anomalous control flags or erratic RST segments do not always stem from malicious exploits; they can also be induced by delayed packet arrivals after a 5G handover, NAT rebinding faults, or middlebox rewriting flaws. The challenge is that their static fields, when isolated, may appear syntactically valid (e.g., the sequence number \texttt{Seq} falls within the valid receive window), yet their presence conflicts with the historical semantics of the active connection.

The LLM uses connection history to identify semantic mismatches and propose handling actions. The original evaluation contains 200 action-labeled records across four scenario groups (50 per group). Table~\ref{tab:comprehensive_slow_path} reports 99.50\% agreement with the action labels in this category. The two RST groups assess history-dependent recognition of suspicious reset segments, while the isolated FIN and SYN+PSH groups are evaluated against their respective labels. Before execution, the Protocol Guard checks the proposed candidate.

\subsubsection{Stateful Triage for Flow-Control Deadlocks}
While the previous scenario primarily evaluates the LLM's ability to drop inconsistent segments, flow-control anomalies require case-by-case handling. Multiple sub-scenarios can share the same surface-level header feature: the peer advertises a zero-window ($\text{Win}=0$). Traditional fast-path state machines or shallow classifiers, when relying mainly on the identical header state, show limited ability to diagnose the underlying causes and may apply a uniform response.

To handle this ambiguity, the slow-path LLM uses historical context to distinguish the five original dataset sub-scenarios and proposes one of their action labels: \texttt{TRIGGER\_PROBE}, \texttt{RST}, or \texttt{DROP}. The 200 labeled records yield 99.00\% action-label agreement in the second part of Table~\ref{tab:comprehensive_slow_path}.

For example, the LLM may propose \texttt{TRIGGER\_PROBE} after interpreting a history of window exhaustion. The guard authorizes a one-byte probe decision only when a trustworthy zero-window update, pending data, an expired probe timer, and a retransmittable outstanding byte are present in the tested branch; otherwise its covered response is to wait. Under the implemented rules, unanswered probes alone do not trigger a reset, and advertised window shrinkage is not automatically treated as a reason to drop a segment~\cite{TCP-RFC}.

\subsubsection{\texorpdfstring{Handling Benign Packets Misrouted by the Classifier}{Handling Benign Packets Misrouted by the Classifier}}
Beyond managing genuine protocol anomalies, a dual-brain architecture should tolerate component-level errors. Section~\ref{sec:4_2} reports a 0.0677\% Classifier false-positive rate on the held-out path-selection set. A classifier false positive can divert a legitimate packet into the slow path; if the LLM then drops it without checking context, the transfer may be disrupted. The slow path must therefore be able to propose \texttt{FORWARD} for benign misrouted traffic.

To test the LLM's action proposal after classifier false positives, we use a counterfactual routing test: 300 benign state-transition records are labeled $p=1$ and forced into the slow path. The LLM uses the packet, state, and history to predict the dataset's \texttt{FORWARD} label. This per-case model test is distinct from connection recovery and from the deterministic protocol-safety fallback in Section~\ref{sec:protocol_guard}. The third part of Table~\ref{tab:comprehensive_slow_path} reports label agreement across seven TCP boundary phases.

The slow-path model matches 99.00\% of the original benign-pass-through labels. Agreement is highest in routine handshake and established-state rows; the LAST-ACK row reaches 98.0\%. These results characterize the model's response to benign traffic misrouted by the classifier.

\subsubsection{Latency--Robustness Trade-off Analysis}
In terms of computational overhead, the average single-packet inference delay of 1033.84~ms for Qwen2.5-7B on a single RTX 5090 GPU is too high for direct deployment on the streaming data plane of a physical network card. However, this latency should be interpreted together with the routing design.

Smart-TCP follows an asymmetric latency--robustness trade-off. In steady-state operation, the front-end Classifier is intended to steer routine packets away from the slow path, so that most frequent data packets are processed by the prototype SLM path (317.79 ms). The approximately one-second computational cost of the 7B large language model is reserved for complex protocol cases, which carry a higher risk of session deadlock or failure. The state synchronization policy in Section~\ref{sec:operational_loop} specifies event ordering during either path's inference. Deployment evaluation will additionally require queueing and timer-response measurements under sustained event arrivals; asynchronous control-plane integration is a future direction.

\subsection{Discussion and Limitations}
\label{sec:limitations}
The current implementation should be interpreted as a prototype-level validation of model-assisted TCP control. The experiments use reconstructed datasets, representative anomaly cases, and simulated data-transfer scenarios to examine the proposed control loop. Future work will integrate Smart-TCP into an operating-system kernel or a high-performance user-space transport stack and evaluate event handling, state synchronization, and system overhead under realistic workloads.

The slow path introduces a latency constraint. In Smart-TCP, the LLM is reserved for rare anomalous, boundary, or fallback cases where an incorrect decision may cause session failure. Future work will investigate model compression and distillation, together with caching and asynchronous control-plane scheduling, to reduce inference latency and computational cost.

The current prototype does not update its models online. Changes in traffic patterns and network conditions may therefore require further adaptation. Future work will examine how to detect such changes, update the models safely, and verify that their decisions remain consistent with protocol constraints.

Finally, Smart-TCP focuses on the feasibility of model-assisted TCP control. Congestion control remains a direction for future work. Extending the framework to congestion-control decisions would enable a subsequent comparison with mature TCP implementations in terms of throughput, fairness, and stability under comparable conditions.

\section{Conclusion}
\label{sec:5}
This paper presents Smart-TCP, an agentic AI-based framework for model-assisted TCP control. Its Dual-Brain architecture routes routine decisions to an SLM and anomalous or boundary cases to an LLM, uses connection state and packet history to inform decisions, delegates arithmetic on sequence and acknowledgment numbers to an ALU, and checks candidates with an RFC-derived Protocol Guard before execution. This division of responsibilities places model reasoning inside a stateful TCP control loop while retaining deterministic arithmetic and protocol checks.

In the evaluated settings, the SLM+ALU fast path achieved 100\% atomic packet accuracy, and Smart-TCP completed all 300 prescribed ideal session windows. The slow-path 7B model achieved 99.14\% action accuracy on 700 labeled anomaly and fallback cases; recovery-action accuracy increased from 4.14\% with packet input alone to 99.14\% with packet, state, and history. In the data-transfer study, 96.5\% of ordinary 32-KiB runs completed at 20\% segment loss with live model inference. Together, these results support the feasibility of the proposed model-assisted control loop within the tested prototype.

\balance
\bibliographystyle{IEEEtran} 

\bibliography{references}

\begin{IEEEbiography}[{\includegraphics[width=1in,height=1.25in,clip,keepaspectratio]{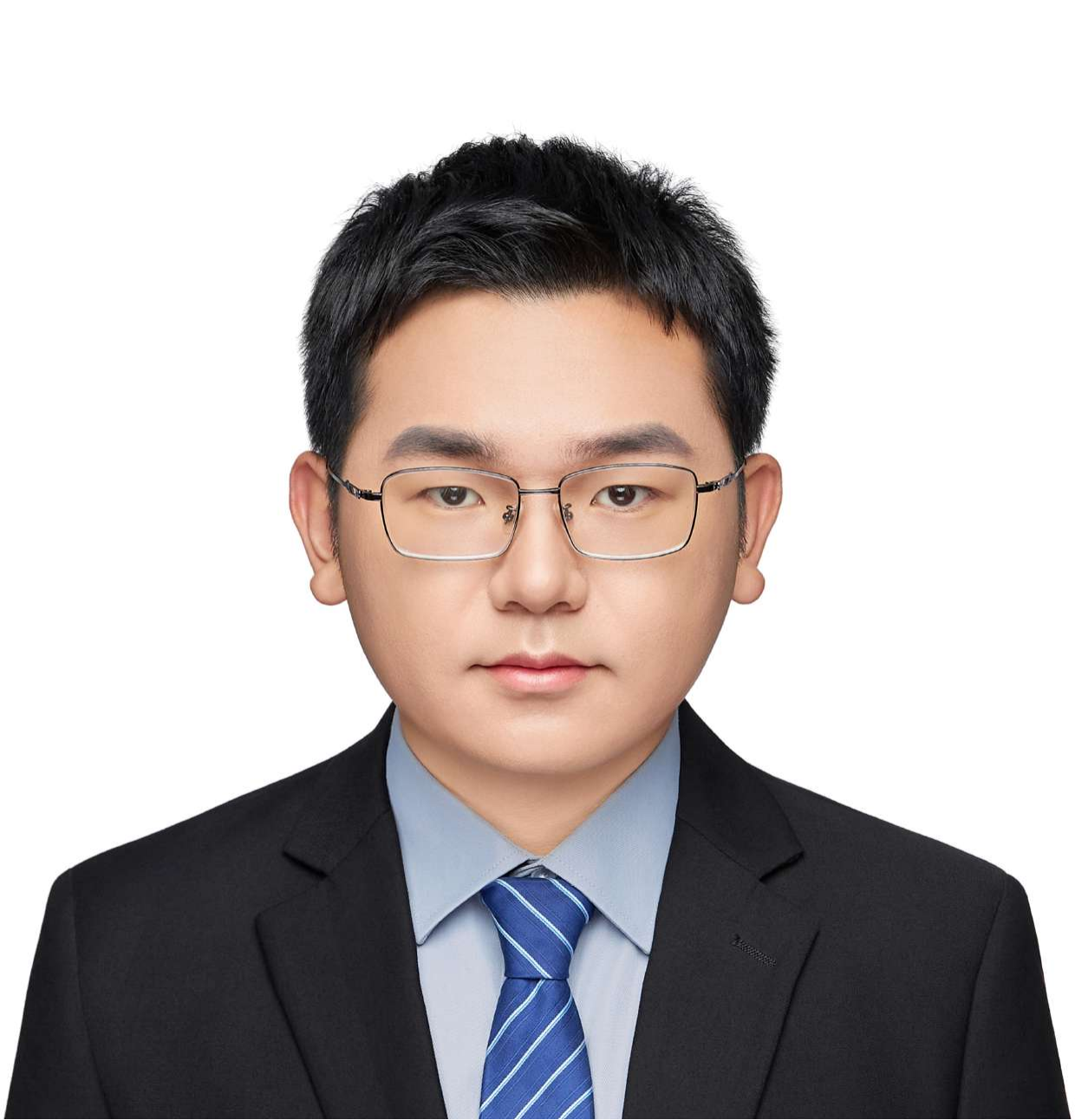}}]{Yule Han}
received the B.S. degree from the University of Electronic Science and Technology of China in 2024. He is currently pursuing the M.S. degree in Information and Communication Engineering, University of Electronic Science and Technology of China, Chengdu, China. His research interests include network protocols, digital twins, and artificial intelligence (AI) for communications.
\end{IEEEbiography}

\begin{IEEEbiography}[{\includegraphics[width=1in,height=1.25in,clip,keepaspectratio]{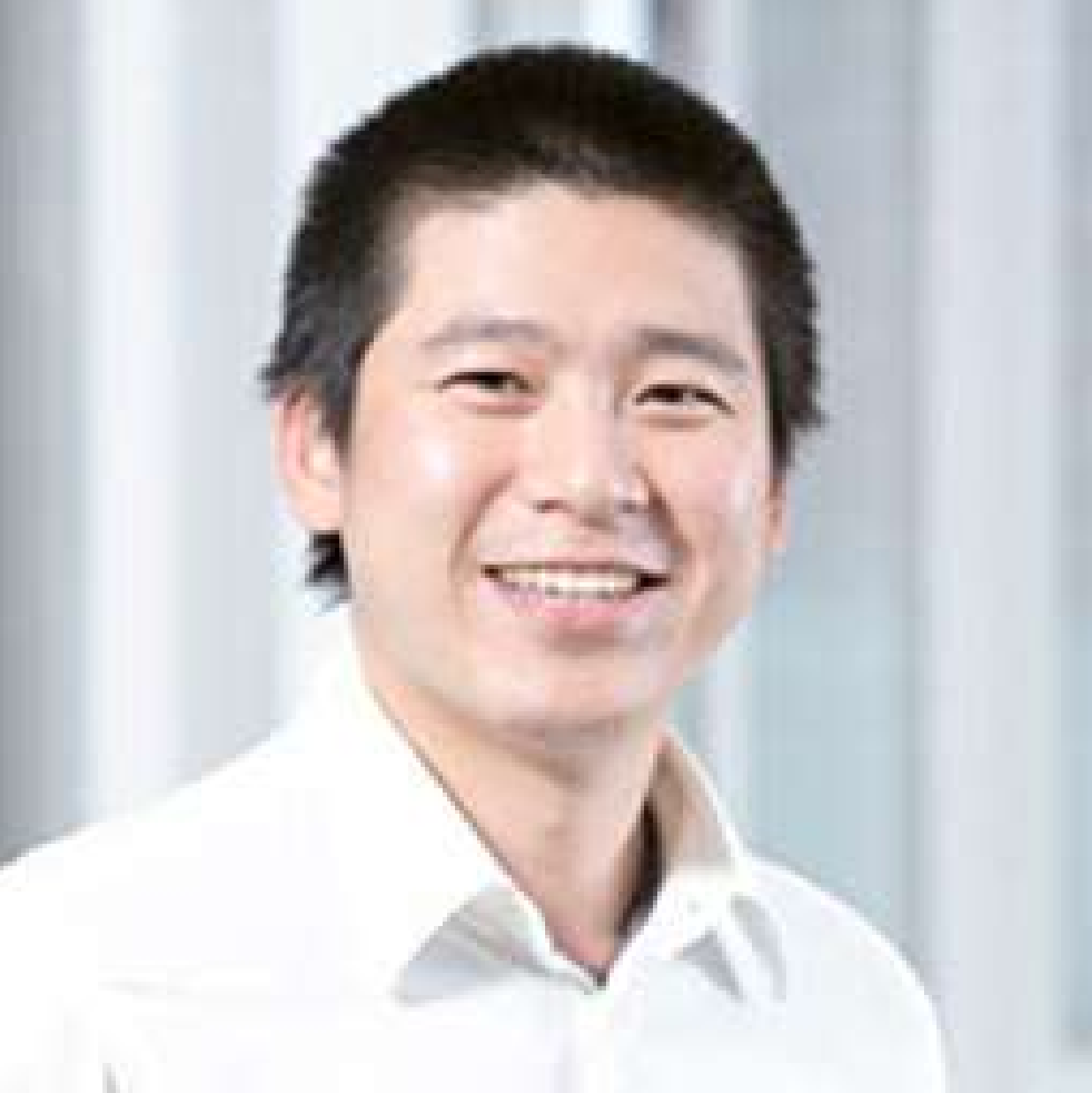}}]{Kezhi Wang}
received the Ph.D. degree from the University of Warwick, U.K. He is currently a Professor in the Department of Computer Science at Brunel University of London, U.K. His research interests include wireless communications, edge computing, and machine learning. He is a Clarivate Highly Cited Researcher.
\end{IEEEbiography}

\begin{IEEEbiography}[{\includegraphics[width=1in,height=1.25in,clip,keepaspectratio]{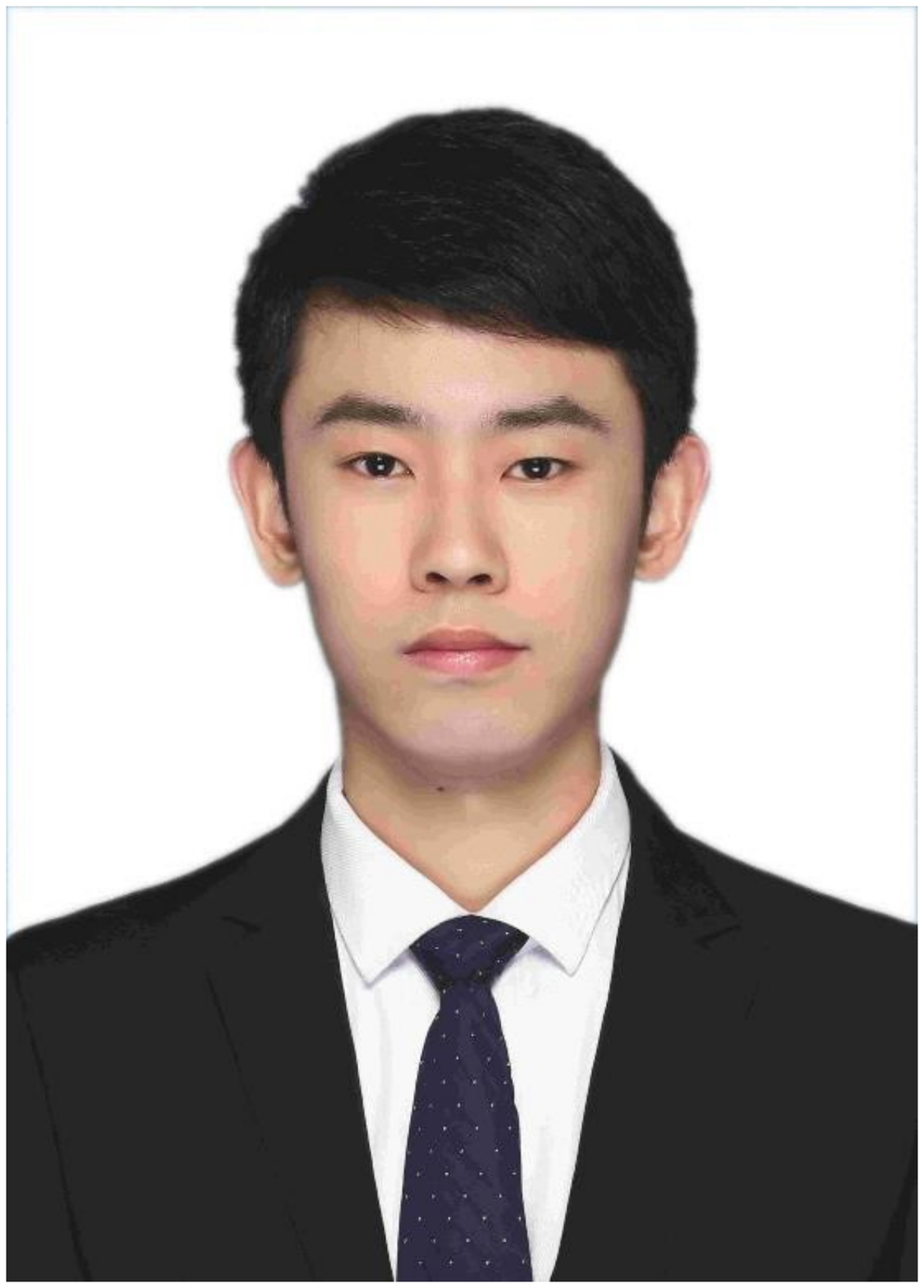}}]{Yizhe Zhao}
(Member, IEEE) received the Ph.D. degree from the School of Information and Communication Engineering, University of Electronic Science and Technology of China (UESTC), in 2021. He has been a Visiting Researcher with the Department of Electrical and Computer Engineering, University of California, Davis, USA. He is currently an Associate Professor with the School of Information and Communication Engineering, UESTC. He is selected in the Young Elite Scientists Sponsorship Program by China Association for Science and Technology (CAST). He is a Senior Member of China Institute of Communications. He was a recipient of the IEEE CSE Best Paper Award in 2023. He serves for China Communications and Journal of Communications and Information Networks (JCIN) as the Youth Editor. He also served as the lead chair and organized workshops on IEEE VTC and Globecom. His research interests include reconfigurable antenna systems, integrated data and energy transfer.
\end{IEEEbiography}

\begin{IEEEbiography}[{\includegraphics[width=1in,height=1.25in,clip,keepaspectratio]{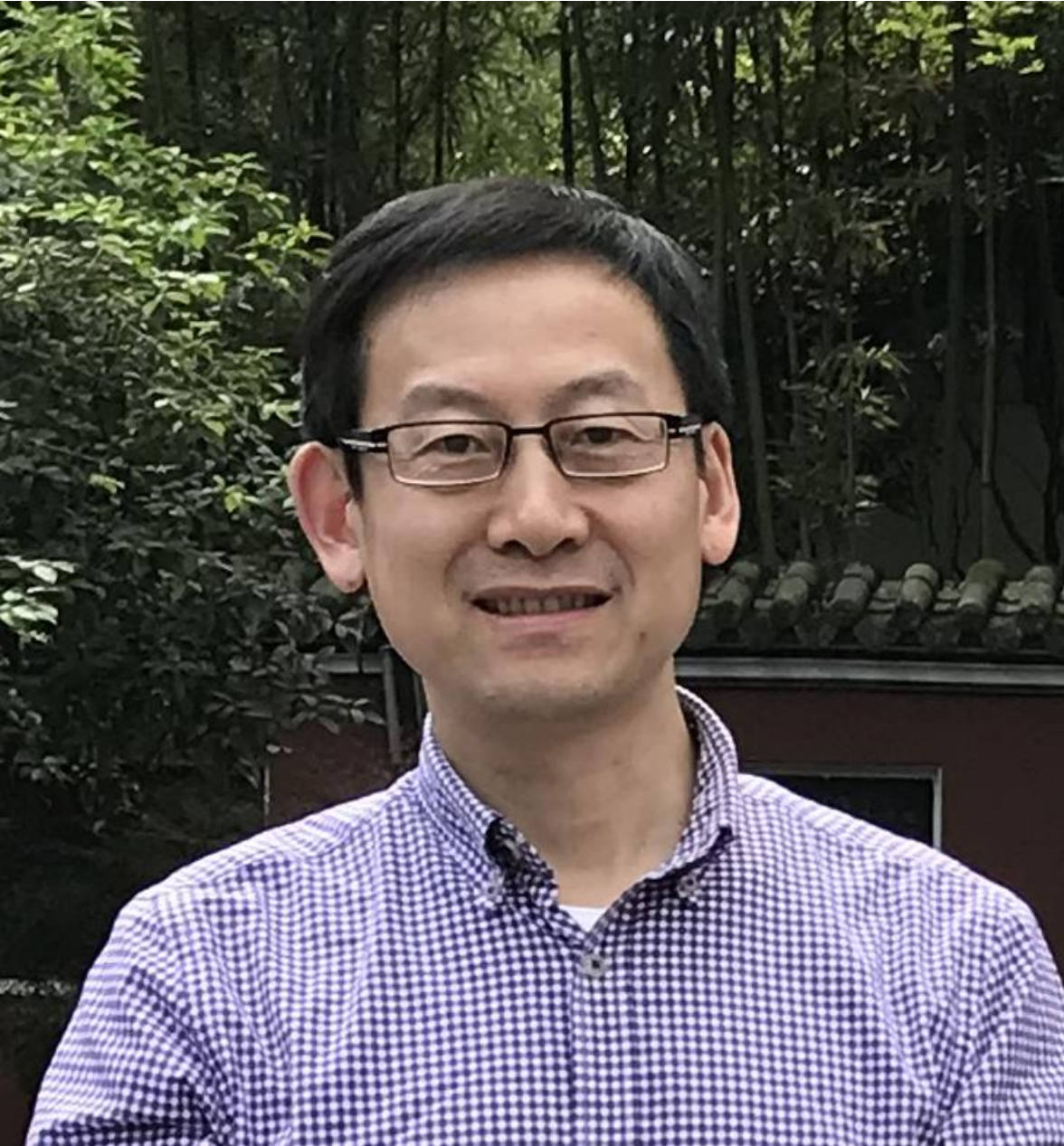}}]{Kun Yang}
received his Ph.D. degree from the Department of Electronic \& Electrical Engineering of University College London (UCL), U.K. He is currently a Chair Professor of Nanjing University and an affiliated professor at the University of Essex. His main research interests include wireless networks and communications, communication-computing cooperation, and AI (artificial intelligence) for wireless. He has published 500+ papers and filed 50 patents. He serves on the editorial boards of a few IEEE journals (e.g., IEEE WCM, TVT, TNB). He is a Deputy Editor-in-Chief of IET Smart Cities Journal. He has been a Judge of the GSMA GLOMO Award at the World Mobile Congress--Barcelona since 2019. He was a Distinguished Lecturer of the IEEE ComSoc, a recipient of the 2024 IET Achievement Medals, and a recipient of the 2024 IEEE CommSoft TC's Technical Achievement Award. He is a Member of Academia Europaea (MAE), a Fellow of IEEE, a Fellow of IET, and a Distinguished Member of ACM.
\end{IEEEbiography}

\end{document}